\documentclass[12pt]{article}

\usepackage[utf8]{inputenc}
\usepackage[makeroom]{cancel}
\usepackage[toc,title]{appendix}
\usepackage{float}
\usepackage{multirow}
\usepackage{color}
\usepackage{caption}
\usepackage{subcaption}
\usepackage{amsmath,amsfonts,amssymb,amsthm,array,relsize}
\usepackage[left=2.5 cm,right=2.5cm,top=3cm,bottom=2.5cm]{geometry}
\theoremstyle{definition}

\usepackage[sort]{natbib} 
\usepackage[multiple]{footmisc}
\usepackage{graphicx} 
\usepackage{breakcites}

\usepackage{hyperref}

\makeatletter
\renewcommand*\@fnsymbol[1]{\the#1}

\def\bbeta{\mbox{\boldmath$\beta$}}

\def\bgamma{\mbox{\boldmath$\gamma$}}

\title{The Tilted Beta Binomial Linear Regression Model: a Bayesian Approach}
\author{Edilberto Cepeda-Cuervo\thanks{Departamento Estad\'istica, Universidad Nacional de Colombia. Email:ecepedac@unal.edu.co}, Mar\'ia Victoria Cifuentes-Amado \footnote{Departamento Matem\'aticas, Universidad Nacional Colombia. Email: mvcifuentesa@unal.edu.co}}

\usepackage{listings}
\begin{document}
\bibliographystyle{apalike}
\date{}
\maketitle

\begin{abstract}
This paper proposes new linear regression models to deal with overdispersed binomial datasets. These new  models, called tilted beta binomial regression models, are defined from the tilted beta binomial distribution, proposed assuming that the parameter of the binomial distribution follows a tilted beta distribution. As a particular case of this regression models, we propose the beta rectangular binomial regression models, defined from the binomial distribution assuming that their parameters follow a beta rectangular distribution. These new linear regression models, defined assuming that the parameters of these new distributions follow regression structures, are fitted applying Bayesian methods and using the OpenBUGS software. The proposed regression models are fitted to an overdispersed binomial dataset of the number of seeds that germinate depending on the type of chosen seed androot.
\end{abstract}


\section{Introduction}
The binomial distribution is normally used to model the number of successes obtained in a finite number of experiments.  However, in these cases, it is often found that the variance of the response variable $Y$ exceeds the theoretical variance of the binomial distribution. This phenomenon, known
as extra-binomial variation (overdispersion), can lead to underestimation errors, lost efficiency of estimates and  underestimation of the variance, wich that can in turn generate incorrect inferences about the regression parameters or the credible intervals \citep*{Williams1982, Cox1983, Collet1991}.

There are several approaches to study overdispersed binomial datasets. \cite{hinde+demetrio1998} categorized the majority of  overdispersed binomial models in two classes: (1) those in which a more general shape for the variance function is assumed, by adding additional parameters; and (2) models in which it is assumed that the parameter of the distribution of the response variable is itself a random variable. In the first class, the double exponential family of distributions allows the researcher to obtain  double binomial models, which allow including a second parameter, which independently from the mean controls for the variance of the response variable and  can be modeled from a subset of some
explanatory variables \citep{Efron86}. In the second class, the beta binomial distribution, results by assuming that the response variable follows a binomial distribution and the probability parameter of the binomial distribution follows a beta distribution. From the parameterization of the beta distribution, in terms of its mean and dispersion parameter \citep{jorgensen1997}, a parameterization of the beta binomial beta distribution in terms of its mean and dispersion parameters  is presented in \cite{Cepeda+Cifuentes2017}.

Despite the versatility of the beta distribution, \cite{hahn2008} proposed the rectangular beta distribution as a combination between the beta distribution and the uniform distribution, to admit heavier tails than that admitted by the beta distribution. After that, \cite{hahn2015} introduced tilted beta distribution, which has as particular cases the beta rectangular and  the beta distributions.

In this article, we generalize the beta binomial regression models for fitting overdispersed binomial count dataset \citep{Cepeda+Cifuentes2017} by introducing the tilted beta binomial linear regression model. For this, the tilted beta
binomial probability is defined by assuming that the parameter of the binomial distribution follows the mean tilted beta distribution.
In addition, the beta rectangular binomial models are presented as particular cases of the new proposed model, by assuming that the parameter of the
binomial distribution has beta rectangular distribution. The proposed models are fitted using Bayesian methods.  Finally, in order to illustrate of the tilted beta binomial model, we fit it to a seed germination count dataset  and compare it with the rectangular beta binomial model and the binomial model, using their DIC values.

This paper is organized as follows. After the introduction, in Section 2, the tilted and the reparameterized tilted beta distributions are presented. In Section 3, the tilted beta binomial distribution is introduced and the rectangular beta binomial distribution is presented as a particular case. In Section 4, the tilted beta binomial linear regression model is defined. Finally, in Section 5, we analyze how the proportion of seeds that germinated on each of 21 dishes, is influenced by the type of seed and root, by fitting a tilted beta binomial linear regression, using the OpenBUGS software. The proposed model performance is compared with the binomial and beta binomial regression models.

\section{The Tilted Beta Distribution}
In different fields there is often a need to model  continuous random
variables that assume values in a bounded interval on a set of explanatory variables.
\cite{cepeda011} proposed the beta regression models, where mean and dispersion parameters
follow regression structures (see also \cite{CepedaGam05}, \cite{CepedaGarrido15}). If the continuous variable $Y$
assumes values in a bounded open interval $(a, b)$, a  beta regression models can be proposed, using the basic transformation
$(y - a)/(b - a)$. However, in order to admit heavier tails than is possible in the beta distribution, \cite{hahn2008} proposed the rectangular beta distribution as a new distribution that, like the beta distribution, has as domain the open interval $(0,1)$. The rectangular beta distribution consists of convex combination between the beta distribution and the uniform distribution $U(0,1)$. Subsequently \cite{hahn2015}, proposed the tilted beta distribution, consisting of a mixture of the beta distribution and the tilted distribution, which has as particular cases the beta rectangular distribution and the beta distribution. This section presents a reparameterization of the tilted beta distribution proposed by \cite{hahn2015}, in terms of the mean and the dispersion parameters of the beta distribution $\mu_b$ and $\phi$, respectively, and the mean of the tilted beta distribution $\mu_t$. The ($\mu_t$,$\mu_b$,$\phi$,$\theta$)-tilted beta binomial distribution results from the convex combination between the tilted reparameterized beta distribution and the binomial distribution.

\subsection{The Tilted Distribution}\label{c_lineal}
A random variable $Y$ follows an inclined distribution with a parameter $\nu$  \citep{hahn2015} if its density is given by:

\begin{equation}\label{densidad_c-lineal}
c(y|\nu)=\left[2\nu-2(2\nu-1)y\right]I_{(0,1)(y)},\quad 0\leq\nu\leq1
\end{equation}
The mean of $Y$,  denoted $\mu_t:=E(Y|\nu)$,  is equal to $\mu_t=(2-\nu)/3$. By reparameterizing (\ref{densidad_c-lineal}) in terms of the mean, the density function is defined by:

\begin{equation}\label{densidad_c-lineal_reparam}
c(y|\mu_t)=\left[3(2\mu_t-1)(2y-1)+1\right]I_{(0,1)(y)},
\end{equation}
where $1/3\leq\mu_t\leq 2/3$, given that the moments, $E_t(Y^n)$, of a random variable $Y$ which follows the density function (\ref{densidad_c-lineal_reparam}) are given by:

\begin{align}
E_t(Y^n)&=\int_0^1 y^n\left[3(2\mu_t-1)(2y-1)+1\right]dy\nonumber\\
&=\frac{3n(2\mu_t-1)+n+2}{(n+1)(n+2)}, \;\; n=1,2,...\label{momentos_c-lineal}
\end{align}
Their variance, $V_t(Y)$,  is given by:
\begin{align}
V_t(Y)&=\frac{6(2\mu_t-1)+4}{12}-\mu_t^2\nonumber
\\&=\mu_t(1-\mu_t)-\frac{1}{6}.\nonumber
\end{align}

\subsection{Reparameterized Tilted Beta Distribution}
The tilted beta distribution was introduced by \cite{hahn2015}, as the convex combination between the tilted distribution and the beta distribution. If this distribution is obtained from the combination of the mean tilted distribution (\ref{densidad_c-lineal_reparam}) and the
mean and the dispersion beta distribution, $Beta(\mu_b, \phi)$, the density function of the tilted beta distribution is given by (\ref{densidad_beta_c-lineal}):

\begin{align}\label{densidad_beta_c-lineal}
f(y|\mu_t,\mu_b,\phi,\theta)=\theta c(y|\mu_t)+(1-\theta)f_{Beta}(y|\mu_{b},\phi)
\end{align}
where $0\leq\theta\leq1$. The notation $Y\sim BI(\mu_t,\mu_b,\phi,\theta)$ is used to denote that $Y$ follows a tilted beta distribution.
Since the $n$-th-moment of $Y$ is given by:

\begin{align}
E(Y^n)&=\int^1_0 y^n\theta c(y|\mu_t)dy+\int^1_0 y^n(1-\theta)f_{Beta}(y|\mu_{b})dy\nonumber\\
&=\theta  E_t(Y^n)+(1-\theta)E_{b}(Y^n)\nonumber\\
&=\theta\frac{3n(2\mu_t-1)+n+2}{(n+1)(n+2)}+(1-\theta)\frac{\Gamma(\mu_{b}\phi+n)\Gamma(\phi)}{\Gamma(\mu_{b}\phi)\Gamma(\phi+n)},\label{momentos_beta_c-lineal}
\end{align}
the mean and the variance of the tilted beta distribution are:

\begin{align}
E\left(Y|\mu_t,\mu_{b},\phi,\theta\right)&=\theta\mu_t+(1-\theta)\mu_t\label{esperanza_beta_c-lineal}\\
V\left(Y|\mu_t,\mu_{b},\phi,\theta\right)&=E(Y^2|\mu_t,\mu_{b},\phi,\theta)-E(Y|\mu_t,\mu_{b},\phi,\theta)^2\nonumber\\
&=\left[\theta E_(Y^2)+(1-\theta)E_b(Y^2)\right]-\left[\theta\mu_t+(1-\theta)\mu_b\right]^2\nonumber\\
&=\theta V_t(Y)+(1-\theta)V_b(Y)+\theta\mu_t^2+(1-\theta)\mu_b^2-\left[\theta\mu_t+(1-\theta)\mu_b\right]^2 \nonumber\\
&=\theta V_t(Y)+(1-\theta)V_b(Y)+\theta(1-\theta)(\mu_t+\mu_b)^2\label{varianza_beta_c-lineal}
\end{align}

The rectangular beta distribution \citep{hahn2008} is a particular case of (\ref{densidad_beta_c-lineal}) when $\mu_t=0.5$ (the slope of the tilted distribution is zero). By replacing this value of $\mu_t=o$ in (\ref{densidad_beta_c-lineal}), the density function of the tilted beta distributions is defined by:
\begin{align}\label{distribucion_beta_rectangular}
f(y|\mu,\phi,\theta)=\theta+(1-\theta)f_{Beta}(y|\mu,\phi).
\end{align}

\section{($\mu_t$,$\mu_b$,$\phi$,$\theta$) - Tilted Beta Binomial Distribution}

Let $Y|p\sim Bin(m,p)$ be a random variable that follows the binomial distribution, where $p$ follows the tilted beta distribution,  $p\sim BI(\mu_t,\mu_b,\phi,\theta)$. Then $Y$ follows a tilted beta binomial distribution with parameters $\mu_t$, $\mu_b$, $\phi$ and $\theta$,
denoted by $Y\sim BIB(\mu_t,\mu_b,\phi,\theta)$. The probability of this distribution is given by:
\begin{align}
f(y|\mu_t,\mu_b,\phi,\theta)&=\int_0^1f_{Bin}(y|m,p)\left[\theta c(p|\mu_t)+(1-\theta)f_{Beta}(p|\mu_b,\phi)\right]dp\nonumber\\
&=\left(\begin{array}{c} m \\ y\end{array}\right)\left[\theta \int_0^1[3(2\mu_t-1)(2p-1)+1] p^y(1-p)^{m-y}dp+\frac{(1-\theta)\Gamma(\phi)}{\Gamma(\mu_b\phi)\Gamma(\phi(1-\mu_b))}\right.\nonumber\\
&*\left.\int_0^1 p^{y+\mu_b\phi-1}(1-p)^{m-y+\phi(1-\mu_b)-1}dp\right]\nonumber\\
&=\theta\left(\begin{array}{c} m \\ y\end{array}\right)\left[(12\mu_t-6)\int_0^1 p^{y+1}(1-p)^{m-y}dp+(-6\mu_t+4)\int_0^1 p^{y}(1-p)^{m-y}dp\right]\nonumber\\
&+(1-\theta)f_{BB(\mu_b,\phi)}(y|\mu_b,\phi)\nonumber\\
&=2\theta\left(\begin{array}{c} m \\ y\end{array}\right)\left[(6\mu_t-3)\frac{y+1}{y+1+m-y+1}+(-3\mu_t+2)\right]B(y+1,m-y+1)\nonumber\\
&+(1-\theta)f_{BB(\mu_b,\phi)}(y|\mu_b,\phi)\nonumber\\
&=2\theta\left(\begin{array}{c} m \\ y\end{array}\right)\left[\frac{y(6\mu_t-3)+m(2-3\mu_t)+1}{m+2}\right]B(y+1,m-y+1)+(1-\theta)f_{BB(\mu_b,\phi)}(y),\label{densidad_BCLBin}
\end{align}
where $B(\cdot,\cdot)$ denotes the beta function and $f_{BB(\mu_b,\phi)}(\cdot)$ denotes the probability function of the beta binomial distribution,  parameterized in terms of the mean and the dispersion parameters.

The behavior of the ($\mu_t,\mu_b,\phi,\theta$)-tilted beta binomial probability function is illustrated in Figure
\ref{Tbetabin}, for different vectors of parameter values:
\begin{figure}[H]
  \centering
   \includegraphics[scale=0.8]{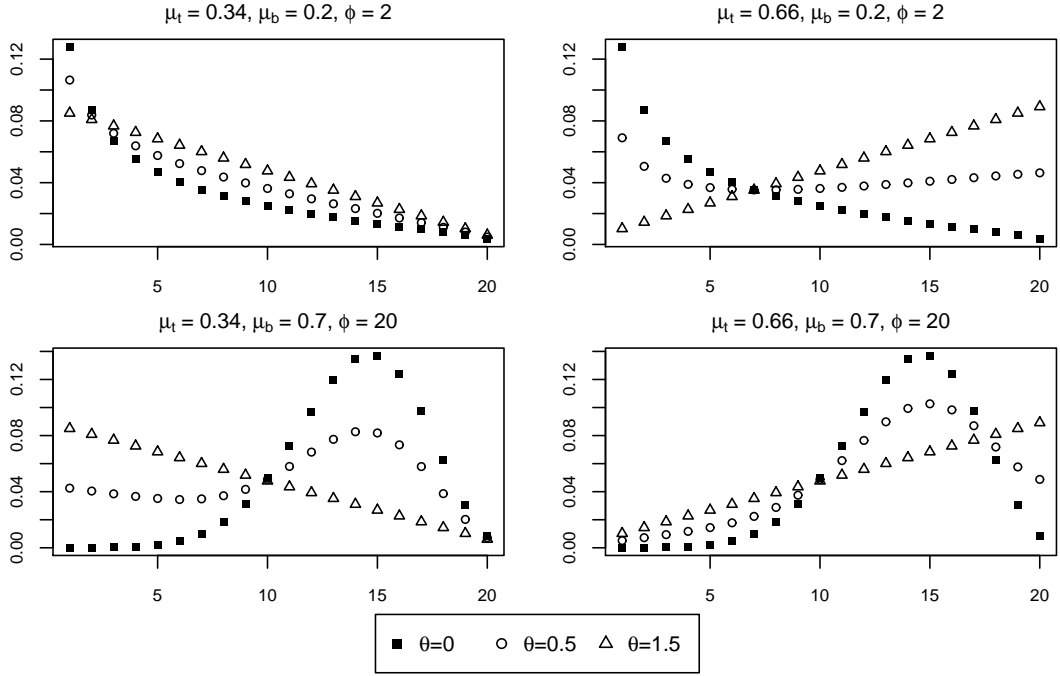}
 \caption{Density function of the ($\mu_t,\mu_b,\phi,\theta$)-tilted beta binomial distribution.}\label{Tbetabin}
\end{figure}

The mean and  variance of a random variable $Y$ that follows the ($\mu_t,\mu_b,\phi,\theta$)-tilted beta binomial probability function are given by:
\begin{small}
\begin{align*}
E(Y)&=E(E(Y|p))=mE(p)\\
&=m\left[\theta\mu_t+(1-\theta)\mu_b\right]\\
V(Y)&=V(E(Y|p))+E(V(Y|p))\\
&=m^2V(p)+mE(p)-mE(p^2))\\
&=m\left\{(m-1)V(p)+E(p)(1-E(p))\right\}\\
&=m\left\{(m-1)\left[\theta V_t+(1-\theta)V_b+\theta(1-\theta)(\mu_t+\mu_b)^2\right]+\left[\theta\mu_t+(1-\theta)\mu_b\right]\left[1-\theta\mu_t+(1-\theta)\mu_b\right]\right\}\\
\end{align*}
\end{small}
where $\mu_b$, $V_b$ denote the mean and  variance of the beta distribution, respectively, and $\mu_t$, $V_t$ denote the mean and  variance of the tilted beta distribution.

\subsection{($\mu_b$,$\phi$,$\theta$)-Beta Rectangular Binomial Distribution}

Let $Y|p\sim Bin(m,p)$ be a random variable that follows the binomial distribution, where $p$ follows the beta rectangular distribution (\ref{distribucion_beta_rectangular}). Thus, $Y$ follows the ($\mu_b$,$\phi$,$\theta$)-beta rectangular binomial distribution.
This density function can be obtained as a particular case of the tilted beta binomial distribution (\ref{densidad_BCLBin}), by replacing $\mu_t$ by $0.5$:

\begin{align}
f(y|\mu_b,\phi,\theta)&=\left(\begin{array}{c} m \\ y\end{array}\right)\theta B(y+1,m-y+1)+(1-\theta)f_{BB(\mu_b,\phi)}(y|\mu_b,\phi)\label{densidad_BRBin}
\end{align}

From the equations of the mean (\ref{esperanza_beta_c-lineal}) and variance (\ref{varianza_beta_c-lineal}) of the tilted beta binomial distribution, setting $\mu_t=0.5$, the mean and variance of the rectangular beta distribution are obtained as:
\begin{small}
\begin{align*}
E(Y)&=m\left[\frac{\theta}{2}+(1-\theta)\mu\right]\\
V(Y)&=(m^2-m)\left[\frac{\mu(1-\mu)}{1+\phi}(1-\theta)(1+\theta(1+\phi)) +\frac{\theta}{12}(4-3\theta)\right]+m\left[\frac{\theta}{2}
+(1-\theta)\mu\right]\left[\frac{2-\theta}{2}-(1-\theta)\mu\right]
\end{align*}
\end{small}

\section{Tilted Beta Binomial Regression Model}
Let $Y\sim BIB(\mu_t,\mu_b,\phi,\theta)$, $i=1,2,\dots,n$, be independent random variables with tilted beta binomial distribution.
Let $\mathbf{x}_i=(x_{i1},...,x_{is})^{T}$, $\mathbf{z}_i=(z_{i1},...,z_{ik})^{T}$ and $\mathbf{w}_i=(w_{i1},...,w_{il})^{T}$ the covariate vectors
of $\mu_b$, $\phi$ and $\theta$ regression structures, and $\bbeta=(\beta_1,...,\beta_s)^{T}$, $\bgamma=(\gamma_1,...,\gamma_k)^{T}$ and $\boldsymbol{\delta}=(\delta_1,...,\delta_l)^{T}$ the respective regression parameter vectors, such that:
\begin{align*}
\mu_{bi}&=\mbox{exp}(\mathbf{x}_i^T\bbeta)/(1+\mbox{exp}(\mathbf{x}_i^T\bbeta))\\
\phi_i&=\mbox{exp}(\mathbf{z}_i^T\boldsymbol{\gamma})\\ \theta_i&=\mbox{exp}(\mathbf{w}_i\boldsymbol{\delta})/(1+\mbox{exp}(\mathbf{w}_i\boldsymbol{\delta}))
\end{align*}
Thus, if  $\mu_t$  is assumed to be constant, the likelihood function of the $BIB(\mu_{t},\mu_{bi},\phi_i,\theta_i)$-regression model is:

\begin{gather*}
l(\mu_t,(\boldsymbol{\beta}^T,\boldsymbol{\gamma}^T,\boldsymbol{\delta}^T)^T)=\frac{\mbox{exp}(\mathbf{w}_i\boldsymbol{\delta})}{1+\mbox{exp}(\mathbf{w}_i\boldsymbol{\delta})}\mathlarger{\sum}_{\{i:y_i\geq0\}}l_{t}(\mu_t|y_i)+\frac{1}{1+\mbox{exp}(\mathbf{w}_i\boldsymbol{\delta})}\mathlarger{\sum}_{\{i:y_i\geq0\}}l_{BB}(\mathbf{x}_i^{T}\boldsymbol{\beta},\mathbf{z}_i^{T}\boldsymbol{\gamma}|y_i)
\end{gather*}
with:
\begin{eqnarray}
l_t(\mu_t|y_i)=log\left\{2\left(\begin{array}{c} m \\ y_i\end{array}\right)\right\}+\mbox{log}\left\{\frac{y(6\mu_t-3)+m(2-3\mu_t)+1}{m+2}\right\}+\mbox{log}\{B(y_i+1,m-y_i+1)\}\nonumber\\
\end{eqnarray}
and
\begin{eqnarray}
l_{BB}(\mathbf{x}_i^{T}\boldsymbol{\beta},\mathbf{z}_i^{T}\boldsymbol{\gamma}|y_i)=\mbox{log}\left\{f_{BB(\mu,\phi)}\left(y_i\left\vert\frac{\mbox{exp}(\mathbf{x}_i^{T}\boldsymbol{\beta}}{1+\mbox{exp}(\mathbf{x}_i^{T}\boldsymbol{\beta}))},\mbox{exp}(\mathbf{z}_i^{T}\boldsymbol{\gamma})\right.\right)\right\},
\nonumber
\end{eqnarray}
where $B(\cdot,\cdot)$ represents the beta function.

In order to define the Bayesian tilted beta binomial regression model, the following a priori distributions are assumed for $\boldsymbol{\beta}$, $\boldsymbol{\gamma}$, $\boldsymbol{\delta}$ and $\mu_t$:

\begin{align*}
\bbeta\sim N(0,\boldsymbol{B})\\
\boldsymbol{\gamma}\sim N(0,\boldsymbol{G})\\
\boldsymbol{\delta}\sim N(0,\boldsymbol{D})\\
\mu_t\sim U(1/3,2/3)
\end{align*}

\section{Seeds Germination Regression Models}
\addtocontents{toc}{\setcounter{tocdepth}{3}}

The dataset analyzed in this section is available in \cite{WinbugsManual2003}openbugsExamples2014 and corresponds to the number of seeds that germinated from an initial quantity arranged   in each of 21 dishes organized according to a  2 by 2 factorial design (2 seed types and 2 root types). These data were initially reported by \cite{seeds+Crowder1978}. The variables involved in the experiment are described below:

\begin{itemize}
\item\textbf{y}: number of seeds germinated in each dish.
\item\textbf{n}: number of seeds initially arranged in each  dish.
\item\textbf{x}$\mathbf{_1}$: seed type (1) if it is $\emph{O. aegyptiaca 75}$ and (2) if it is $\emph{O. aegyptica 73}$.
\item\textbf{x}$\mathbf{_2}$: root type (1) if it is bean and (2) if it is cucumber.
\end{itemize}

In this experiment, there are 21 observations (21 dishes). Since the variable $Y$ counts the number of germinated seeds in each dish, this variable can be modeled by a linear regression TBB($\mu_t$,$\mu_b$,$\phi$,$\theta$) model, which includes all the explanatory variables in each of the regression structures.  After the process of eliminating the explanatory variables, the best model (the model with smallest DIC value) has the following regression structures:

\begin{align*}
\mbox{logit}(\mu_{ib})&=c1+c2*\textbf{x}\mathbf{_1}+c3*\textbf{x}\mathbf{_2}\\
\mbox{log}(\phi_i)&=a1+a2*\textbf{x}\mathbf{_2}\\
\mbox{logit}(\theta_i)&=b1\\
\mu_t&\sim U(1/3,2/3),
\end{align*}
where $i=1,\dots,21$. The TBB($\mu_t$,$\mu_b$,$\phi$,$\theta$) model was fitted to the data using OpenBUGS, a free program used for Bayesian regression based on the Gibbs algorithm  \citep{WinbugsManual2003}. The posterior parameter inferences  obtained from a sample of size 100000, burn-in of the first 10000, and taking one sample every 10 iterations to reduce autocorrelation, are summarized in Table \ref{results_BCLB}. The DIC value of this model is 121.9.

\begin{table}[H]
\centering
\begin{tabular}{|c|c c c c|}
\hline
Parameter & Mean & S.D.& 95\% Cred. Interval  & M.C. Error\\
\hline\hline
$a1$ &	1.398 & 2.141 & (-2.781,5.769) & 0.0268\\
$a2$ &	2.285 &	1.839 & (-0.7843,6.42) & 0.0218\\
$b1$ &	-3.649 & 1.826 & (-7.927,-0.7198) & 0.0223\\
$c1$ &	-0.9479 & 0.5086 & (-1.892,0.1485) & 0.0059\\
$c2$ &	-0.4403 & 0.2352 & (-0.9218,-0.0081) & 0.0026\\
$c3$ &	1.036 & 0.238 & (0.5653,1.51) & 0.02819\\
$\mu_c$ & 0.4941 & 0.0956 & (0.3411,0.6569) & 0.0011\\
$s.d.$ & 116.9 & 3.543 & (112.1,125.6) & 0.0437
\\		
\hline
\end{tabular}
\captionof{table}{Posterior parameter estimates of  $TBB(\mu_t,\mu_b,\alpha,\theta)$ model}
\label{results_BCLB}
\end{table}

\begin{figure}[H]
 \centering
  \begin{subfigure}{0.45\textwidth}
  \centering
   \includegraphics[scale=0.4]{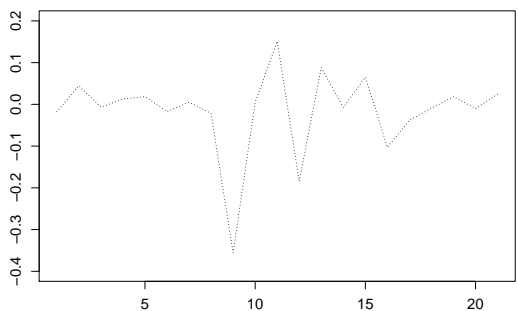}
  \caption{Residuals versus iterations}
  \end{subfigure}
  \hspace{0.05\textwidth}
  \begin{subfigure}{0.45\textwidth}
  \centering
   \includegraphics[scale=0.4]{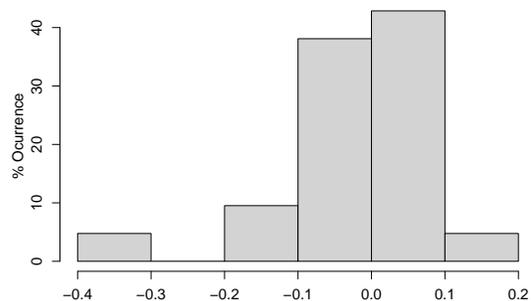}
  \caption{Residual histogram}
  \end{subfigure}
 \caption{Pearson residuals – Seed germination}
 \label{residuales_BCLBin_seeds}
\end{figure}

In Table \ref{results_BCLB} the M.C. error denotes an estimation of the standard Monte Carlo error, wich measures the distance between the posterior estimation of the mean and the mean of the posterior distribution, which is expected to converge to zero when the number of iterations goes to infinity. The Monte Carlo error estimates obtained using the OpenBugs software, close to zero for all the regression parameters, is given by $Error MC\approx DE/\sqrt{Iterations}$ \citep{flegal2008montecarlo}. The  DIC value of this model is 121.9. According to Figure \ref{residuales_BCLBin_seeds}, Pearson's residuals are close to zero, taking values between -0.4 and 0.2, and have no  tendency through the iterations.

\subsection{Chain convergence in the tilted beta binomial model}

In the parameter estimation process, three posterior samples were generated beginning from different starting values. In all  chains,  the autocorrelation is close to zero for a lag greater than or equal than 10, and a burn-in bigger than 10000.

\begin{figure}[H]
  \centering
   \includegraphics[scale=0.5]{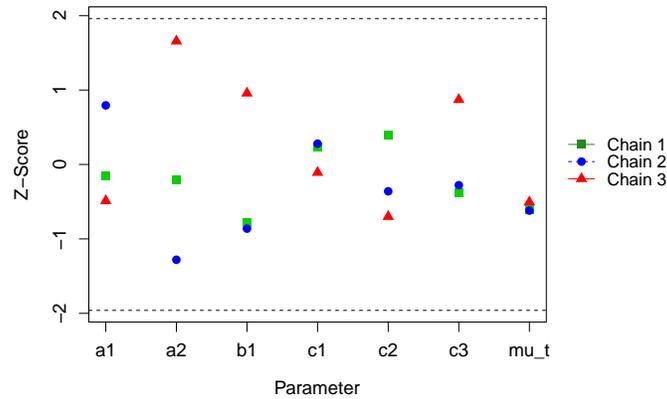}
 \caption{Geweke convergence diagnostic - Seed germination}
 \label{geweke_plot_semillas}
\end{figure}

\begin{figure}[H]
  \centering
   \includegraphics[scale=0.8]{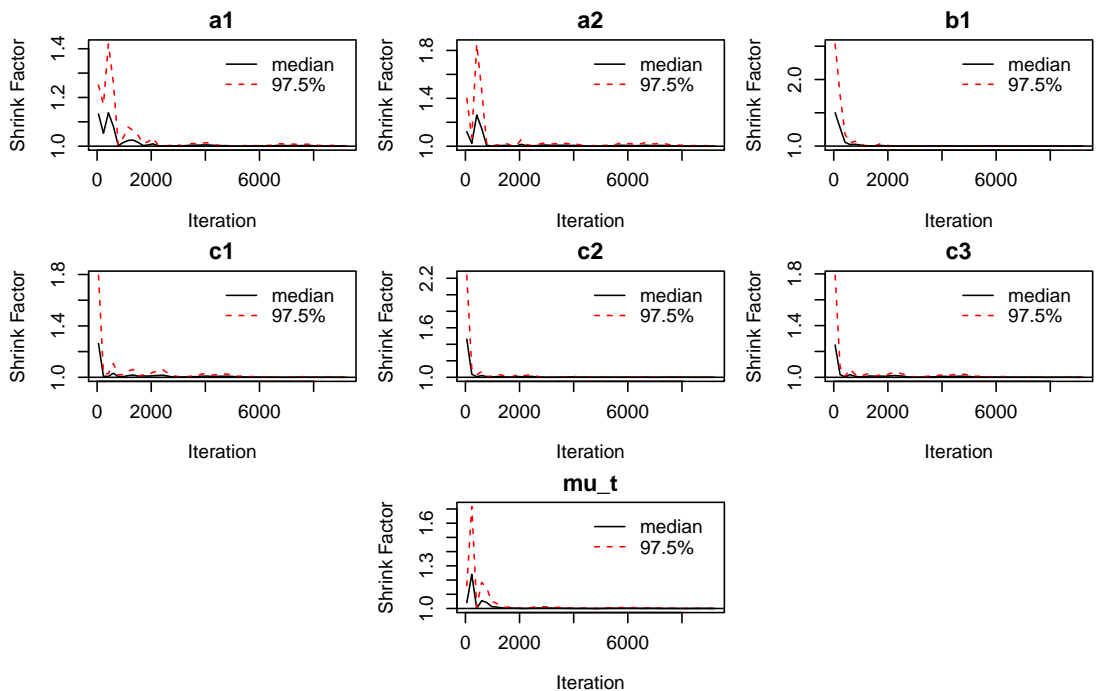}
 \caption{Brooks-Gelman and Rubin convergence diagnostic - Seeds germination}
 \label{diag_gelman+rubin_seeds}
\end{figure}

To check the convergence of the chains, two convergence diagnoses were applied: the Geweke diagnostic \citep{geweke1992} and the Brooks-Gelman-Rubin convergence diagnostic \citep{brooks+gelman1998}. The Geweke-Brooks plot for the chains of the regression parameters can be observed in Figure \ref{geweke_plot_semillas}, where the value of the $Z$ statistic versus the number of iterations is plotted to determine the burn-in of the chains. This figure shows that the statistic remains within the acceptance zone for a period of burn-in equal to zero. The second method applied is known as the Brooks-Gelman and Rubin convergence diagnostic. It was proposed by \cite{brooks+gelman1998} and compares within-chain and between-chain variances through the estimation of the statistic of scale reduction $R$.
Values of $R$ well above 1 indicate that the chains have not converged. Figure \ref{diag_gelman+rubin_seeds} shows that for the regression parameters of this example, the R factor is very close to 1 after the 1000 iterations.

%

\subsection{Models comparison}
In order to determine the performance of the proposed model, the following models also were fitted to the seed germination dataset: binomial $Bin(n,p$), beta binomial $BB(\mu,\phi$) and beta rectangular binomial $BRB(\mu,\phi,\theta$). The deviance  and the deviance information Criterion (DIC) for each of these models are given in Table \ref{comparison_BCLB}, which  shows that the lowest average of the deviance and the lowest DIC value correspond to the tilted beta binomial and the beta rectangular binomial models, where the first one presents the lowest DIC value  and therefore is the best model.
\begin{table}[H]
\centering
\begin{tabular}{|l|l| llll|}
\hline
\multirow{2}{*}{Model} & \multirow{2}{*} {DIC} & \multicolumn{4}{c|}{Deviance}\\
\cline{3-6}
 & & Mean & S.D. & Cred. Interval 95\% & Median\\
\hline\hline
Bin($n$,$p$)& 174.6 & 172.8  & 1.827 & (172.8,179.5) & 174.1\\
BB($mu$,$\phi$)& 154.6 & 149.3  & 2.87 & (145.8,156.6) & 148.6\\
BRB($\mu,\phi,\theta$)& 123.4 &  116.9 & 3.59 & (112.2,125.7) & 116.3\\
TBB($\mu_t,\mu_b,\phi,\theta$)&121.9 &116.7 & 3.54 & (112.1,125.2) & 116\\
\hline
\end{tabular}
\captionof{table}{Statistics for model comparison - Seed germination}
\label{comparison_BCLB}
\end{table}

\section{Conclusion}

In this paper two new distributions are proposed: the tilted beta binomial distribution and the beta rectangular binomial distribution.
From these distributions, assuming that their parameters follow regression structures, new overdispersion regression models for count data are proposed: the tilted beta binomial regression model and the beta rectangular binomial regression model. These models are fitted using Bayesian methods, and in the application,  show better performance  than the beta binomial regression models for statistical analysis of the seed germination dataset.

Given that  the tilted beta distribution is   flexible  and  allows considering varying amounts with greater likelihoods than the beta distribution in the  extreme tail-area events, it permits accommodating different relative likelihoods of high versus low extreme tail-area events.
Thus, the proposed tilted beta binomial regression model which defines a more general overdispersion regression
model than the beta binomial regression model, allows considering count events with  high or low likelihood of occurrence and
better estimation of the regression  parameters, credibility (or confidence) intervals and statistical inferences in the analysis of binomial-type overdispersion data.
\bibliographystyle{jtbnew}
\bibliography{bibliografiaBetaBinC}

\begin{appendices}
\section{OpenBUGS code for the TBB regression model}\label{anexo}
\begin{lstlisting}
model{		
#Likelihood:
 for( i in 1 : N ){
   zeros[i]<-0
   zeros[i]~dloglik(loglike[i])
   loglike[i]<-log(theta[i]*fd_b[i]+(1-theta[i])*fd_BB[i])
 #fd_b: beta function part
 #fd_BB: loglikelihood for beta binomial part
 fd_b[i]<-exp(log(2)+logfact(n[i])-logfact(n[i]-y[i])-logfact(y[i])
  +loggam(y[i]+1)+loggam(n[i]-y[i]+1)-loggam(y[i]+1+n[i]-y[i]+1)
  +log(y[i]*(6*mu_c-3)+n[i]*(2-3*mu_c)+1)-log(n[i]+2))	
 fd_BB[i]<-exp(logfact(n[i])-logfact(n[i]-y[i])-logfact(y[i])
 +loggam(phi[i])+  loggam(y[i]+mu[i]*phi[i])+loggam(n[i]-y[i]
 +phi[i]*(1-mu[i])))-loggam(mu[i]*phi[i])
  -loggam(phi[i]*(1-mu[i]))-loggam(n[i]+phi[i]))
 log(phi[i])<-a[1]+a[2]*(x1[i]+1)
 mu[i]<-max(1.0E-6,mu2[i])
 logit(mu2[i])<-c[1]+c[2]*(x1[i]+1)
 logit(theta[i])<-b[1]+b[2]*(x2[i]+1)
 }
#Priors:
 mu_t~dunif(0.333333,0.666666)
 for (i in 1:2){ a[i] ~ dnorm(0,0.1)}
 for (i in 1:2){ b[i] ~ dnorm(0,0.1)}
 for (i in 1:2){ c[i] ~ dnorm(0,0.1)}
}
\end{lstlisting}

\end{appendices}

\end{document}